\documentclass[aps,twocolumn,prl,fixfloat]{revtex4-2}
\usepackage{graphicx,amsmath,amsfonts,bm, color}

\usepackage{xcolor,hyperref}
\hypersetup{
   colorlinks,
   linkcolor={blue!50!black},
   citecolor={blue!50!black},
   urlcolor={blue!80!black}
}
\usepackage{url} 
\usepackage{lineno}
\usepackage{soul}

\usepackage{enumitem}

\DeclareMathAlphabet{\mathitbf}{OML}{cmm}{b}{it}

\begin{document}

\title{Northbound Lagrangian Pathways of the Mediterranean Outflow Water and the Mechanism of Time-Dependent Chaotic Advection}
\author{Ori Saporta-Katz$^{1,*}$, Nadav Mantel$^2$, Rotem Liran$^2$, Vered Rom-Kedar$^1$, Hezi Gildor$^2$}

\affiliation{$^1$ Department of Computer Science \& Applied Mathematics, Weizmann Institute of Science, Rehovot 7610001, Israel \\ 
$^2$ The Fredy \& Nadin Herrman Institute of Earth Sciences, 
The Hebrew University, The Edmond J. Safra Campus - Givat Ram, Jerusalem 9190401, Israel \\
$^*$ Corresponding author: Ori Saporta-Katz, ori.katz4@mail.huji.ac.il}

\begin{abstract}
The Mediterranean Sea releases approximately 1Sv of water into the North Atlantic through the Gibraltar Straits, forming the saline Mediterranean Outflow Water (MOW). Its impact on large-scale flow and specifically its northbound Lagrangian pathways are widely debated, yet a comprehensive overview of MOW pathways over recent decades is lacking. We calculate and analyze synthetic Lagrangian trajectories in 1980-2020 reanalysis velocity data. 16\% of the MOW follow a direct northbound path to the sub-polar gyre, reaching a 1000m depth crossing window at the southern tip of Rockall Ridge in about 10 years.
Surprisingly, time-dependent chaotic advection, not steady currents, drives over half of the northbound transport. 
Our results suggest a potential 15-20yr predictability in the direct northbound transport, which points to an upcoming decrease of MOW northbound transport in the next couple of decades.
Additionally, monthly variability appears more significant than inter-annual variability in mixing and spreading the MOW.
\end{abstract}

\maketitle

\section{Introduction}

\begin{figure}
 \noindent\includegraphics[width=0.55\textwidth]{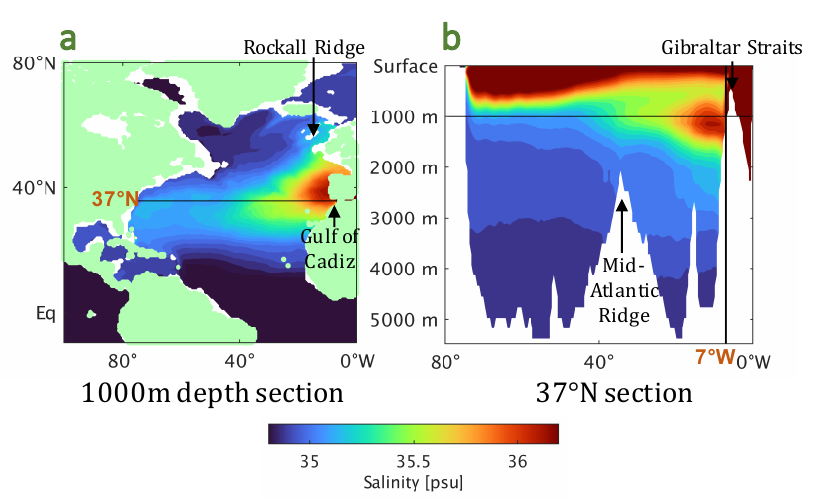}
\caption{
Mediterranean Outflow Water (MOW) in the SODA3.4.2 reanalysis data for climatological annual data averaged 1980-2020. (a) Salinity at 1000m (b) Salinity at 37°N latitude.}
\label{fig:MOWtongue}
\end{figure}

The mid-depth salinity and temperature fields of the North Atlantic Ocean contain a distinct high-salinity, high-temperature tongue originating from the Mediterranean Sea (Fig.~\ref{fig:MOWtongue}). This is a signature of the Mediterranean Outflow Water (MOW), which exits the Straits of Gibraltar into the Gulf of Cadiz as a 0.85 Sv inverse-estuary flow with an average salinity of about $38.4$ psu at a depth of 300-500 meter \cite{BaringerPrice1997, iorga1999signatures, sammartino2015ten}. Upon its entrance into the Gulf of Cadiz (GoC), the relatively salty and dense water sinks to 500-1500 meters entraining the locally fresher and cooler North Atlantic Central Water, and exits the GoC upon crossing the Cape of Vicente as a relatively salty plume of approximately 1 Sv and salinity between $36.3-37$ psu. Beyond this point, the MOW begins spreading in the North Atlantic Ocean via various pathways
\cite{iorga1999signatures, bashmachnikov2015properties, gasser2017tracking, sanchez2017mediterranean}.

The significant input of salinity into the North Atlantic Ocean is thought to have an important effect on the strength and stability of the Atlantic Meridional Overturning Circulation in current and past climates by contributing to the salinity preconditioning of polar waters for formation of the North Atlantic Deep Water (NADW) \cite{Reid1979, Rahmstorf1998, chan2003effects, voelker2006mediterranean, calmanti2006north, rogerson2010enhanced, Ivanovic2014, ferreira2018atlantic}.
The specific pathways taken by the MOW have been of interest and controversy at least since \cite{Reid1978, Reid1979}, who used temperature, salinity, oxygen, and silica data from 1957-1971 to establish hydrographic evidence of a direct northbound pathway of MOW, extending from the Gulf of Cadiz well past Porcupine Bank at 53°N via a mid-depth eastern boundary current.
Since then, our understanding of these pathways has evolved by studies of hydrographic data, actual, and virtual drifters \cite{iorga1999signatures, mccartney2001origin, JiaEtAl2007, SalaEtAl2013, LozierStewart2008, burkholder2011middepth}.
Several works found no evidence that a MOW core exists past Porcupine Bank \cite{iorga1999signatures, mccartney2001origin}. To show this, \cite{iorga1999signatures} used climatological mean fields from 1904-1990, and \cite{mccartney2001origin} used hydrographic methods on data from the 1980s.

In an attempt to reconcile the conflicting evidence, \cite{LozierStewart2008} suggested a temporal variability of flow fields, perhaps due to the North Atlantic Oscillation (NAO), that results in an east-west shift of the eastern limb of the sub-polar gyre (SPG). Using historical hydrographical data and salinity anomalies, they showed the temporal variability of the eastern limb of the SPG as well as a northward penetration of a MOW core between 1000-1500 m upon shrinking of the eastern limb. \cite{chaudhuri2011contrasting}, using a basin-scale model of the North Atlantic, similarly used salinity anomalies to show a MOW core at intermediate depths reaching northward of Porcupine Bank in an extreme NAO low year (1996). \cite{frazao2021mediterranean}, with an array of ARGO floats and CTD data in the Northeast Atlantic between 1981-2018, showed periods of warming (cooling) and salinification (freshening) of the area lagging a negative (positive) NAO index indicating a northward (westward) MOW pathway.
\cite{burkholder2011middepth} were the first to use Lagrangian trajectories (both floats and synthetic) to try and map the specific MOW pathways. Instead of using an atmospheric index (the NAO), \cite{burkholder2011middepth} opted to use an ocean-based measurement since it better depicts the gradual changes in the subpolar gyre, called the gyre index (hereinafter SPG index) using empirical orthogonal function analysis of sea surface height provided by \cite{hakkinen2004decline}. While there were no clear specific pathways from the eastern North Atlantic to the Rockall Trough region, they found that two broadly defined pathways influenced by the SPG index can result in a greater amount of salty MOW-influenced eastern North Atlantic waters reaching the Rockall Trough. It is important to note that different SPG indices may arrive at different results \cite{koul2020unraveling}. \cite{foukal2017assessing} looked at decadal SPG indexes using EOF analysis and defined gyre size using the largest closed contour of each monthly SSH field, which revealed oscillations that do not show an impact of the SPG on salinity anomalies, while  \cite{koul2020unraveling} showed that a density based index captures both salinity anomalies and SPG strength and size.

 \begin{figure*}
 \noindent\includegraphics[width=0.9\textwidth]{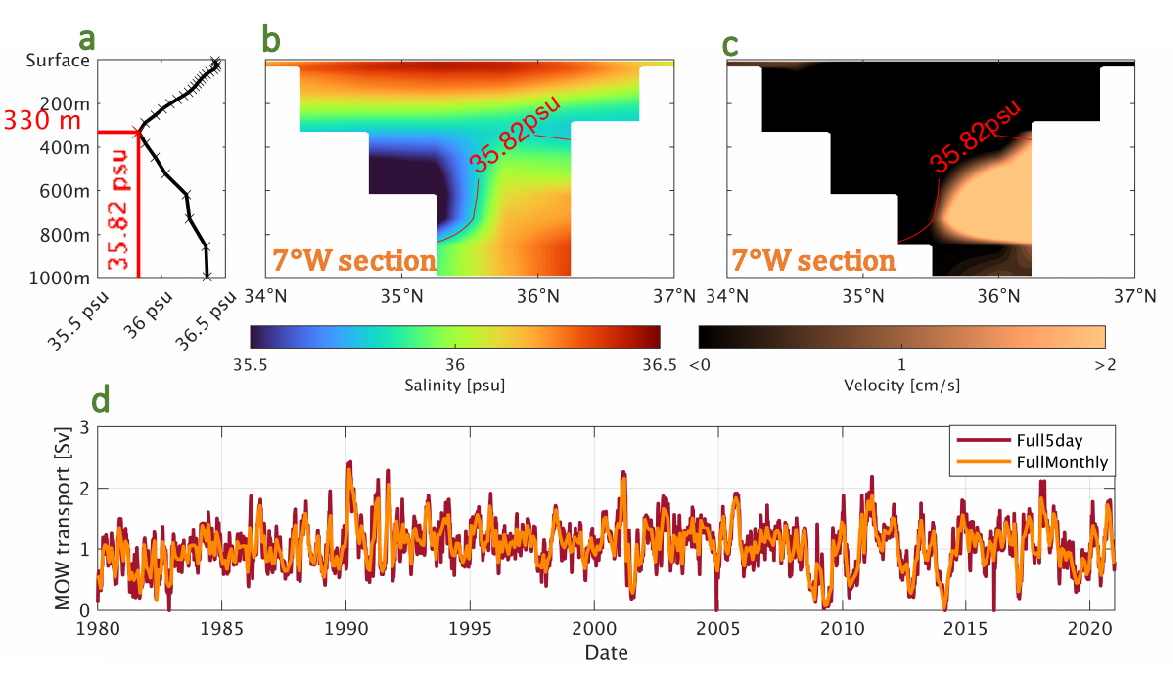}
\caption{
Mediterranean Outflow Water (MOW) in the SODA3.4.2 reanalysis data for climatological annual data averaged 1980-2020. (a) Maximum salinity per depth at 7°W; MOW is defined as the area with a westward velocity and salinity above 35.82 psu at a depth below 330 m, see main text for details.
(b) Salinity at 7°W.
(c) Velocity at 7°W.
(d) MOW transport in Sverdrup for 5-day and monthly SODA data.}
\label{fig:whoisMOW}
\end{figure*}

The time-dependent, 3D nature of oceanic flow over decadal timescales  may result in significant chaotic advection of Lagrangian trajectories, that is accountable for some degree of the oceanic transport and mixing due to stretching of material lines \cite{Aref1984, Ottino1990, YangLiu1994, KoshelPrants2006, ArefEtAl2017}.
The concept of chaos is asymptotic; in finite-time studies, oceanic chaotic advection refers to the divergence of trajectories and Lagrangian mixing on a predetermined timescale \cite{HallerPoje1998, haller2002lagrangian, hadjighasem2017critical}. The associated coherent structures of transport, as exposed in finite times, may reveal the backbone hyperbolic structure of the flow.
While a 3D steady incompressible flow is sufficient for chaotic advection (as opposed to a 2D incompressible steady flow), a special feature of time-dependent flows, both in 2D and 3D, is that streamlines do not equal material lines (Lagrangian trajectories), allowing what we denote ``time-dependent chaotic advection": a transport mechanism that moves a tracer from point A to point B in a predefined timeframe despite there being no streamlines from point A to point B, i.e. no direct flow at any snapshot in time. A well-known 2D example of this phenomenon is the oscillating double-gyre example \cite{Aref1984, YangLiu1994}, where the steady flow has a separatrix that separates the two gyres whereas the unsteady flow has a chaotic transport mechanism between the regions. These studies illustrate the importance of considering time-dependent dynamics to track transport. 
Roughly, this time-dependent mechanism is created in
a flow with several instantaneous saddles (hyperbolic structures) that split nearby trajectories to different directions. If the splitting surfaces themselves are not stationary,  a tracer can ``jump" between cycles and reach places it could not reach in the predefined timescale (or, perhaps, ever).
In 2D, it was shown that
this mechanism is essential 
(due to the integrability of the steady trajectories):
 an average flow plus a reasonable isotropic diffusion will not recreate large-scale time-dependent transport \cite{CarlsonEtAl2010}. 

In this work, we produce and analyze synthetic Lagrangian trajectories of the MOW in realanalysis velocity data to provide a comprehensive analysis of transport statistics and associated timescales. We examine the effect of seasonal, annual, and interannual variations on pathways of the MOW in the entire North Atlantic basin in the past four decades.
To this aim, we compare virtual trajectories  released at the GoC and advected by several types of oceanic models. Their analysis allows us to
evaluate the effect of transient eddies, seasonal, annual, and interannual variability, and the effect of time-dependent chaotic advection on the northbound transport and overall mixing of the MOW in the North Atlantic.

\subsection*{Methods}

Virtual passive tracers released from the Gulf of Cadiz at $7^\circ W$ are advected in the North Atlantic Ocean using velocity fields from the SODA3.4.2 reanalysis spanning 1980-2020 \cite{CartonChepurinChen2018soda3}. 
The SODA3.4.2 reanalysis uses the MOM4p1 code, that solves the Boussinesq hydrostatic  primitive equations \cite{delworth2012simulated, griffies2009elements, griffies2016ocean}. It is an eddy-permitting global ocean-sea ice model with a $1/4$°$\times1/4$° horizontal resolution and 50 vertical levels with a finer resolution towards the surface, forced at the surface by the ERA-interim near-surface atmospheric variables \cite{dee2011era}.
The velocity fields we use for the Lagrangian trajectory integration have been regridded by the SODA team onto a $1/2$°$\times1/2$° centered horizontal grid, and have either a 5-day or a monthly averaged temporal resolution.
The tracking is performed using the Patato toolbox \cite{fredj2016particle}, with linear spatial and temporal interpolators, and an additional freeslip condition on the boundaries that follows the scheme presented in the OceanParcels toolbox tutorials (OceanParcels.com,  \cite{delandmeter2019parcels}). 

We perform four distinct types of experiments, each corresponding to a different type of oceanographic model:
\begin{enumerate}
\item Full5day: 12$\times$20 monthly releases in the full reanalysis 5-day data, from January 1st, 1980 to December 1st, 1999, denoted Full5day\_\textit{\#year}\_\textit{\#month}.
\item FullMonthly: 12$\times$20 monthly releases in the full reanalysis monthly-averaged data, from January 1st, 1980 to December 1st, 1999, denoted FullMonthly\_\textit{\#year}\_\textit{\#month}.
\item RepeatYear: 12$\times$20 monthly releases in yearly-periodic velocity fields, denoted RepeatYear\_\textit{\#year}\_\textit{\#month}. For example, RepeatYear\_n\_m follows trajectories released in year n, month m, into a yearly-periodic velocity field that repeats year n for 20 times.
\item RepeatMonth: 12$\times$20 monthly releases in steady velocity fields that consist of the release month only, denoted RepeatMonth\_\textit{\#year}\_\textit{\#month}. For example, RepeatMonth\_n\_m follows trajectories released in year n, month m, into a steady velocity field that repeats year n, month m, for 12$\times$20 times.
\end{enumerate}
The first and second models are the closest to the real oceanic flow. The first has the maximal possible temporal-spatial resolution and includes inter-annual variability, and the second model is very similar to it, with a slight reduction in the time resolution. The third model corresponds to  time-periodic velocity fields that have the same spatial resolution and kinetic energy as in the second mode. The fourth model is steady, with the same spatial resolution and kinetic energy as the second model. The traditional climatological models suffer, due to the averaging, from a low kinetic energy content, hence they are not included here, see App. A.

 \begin{figure*}[t]
 \noindent\includegraphics[width=\textwidth]{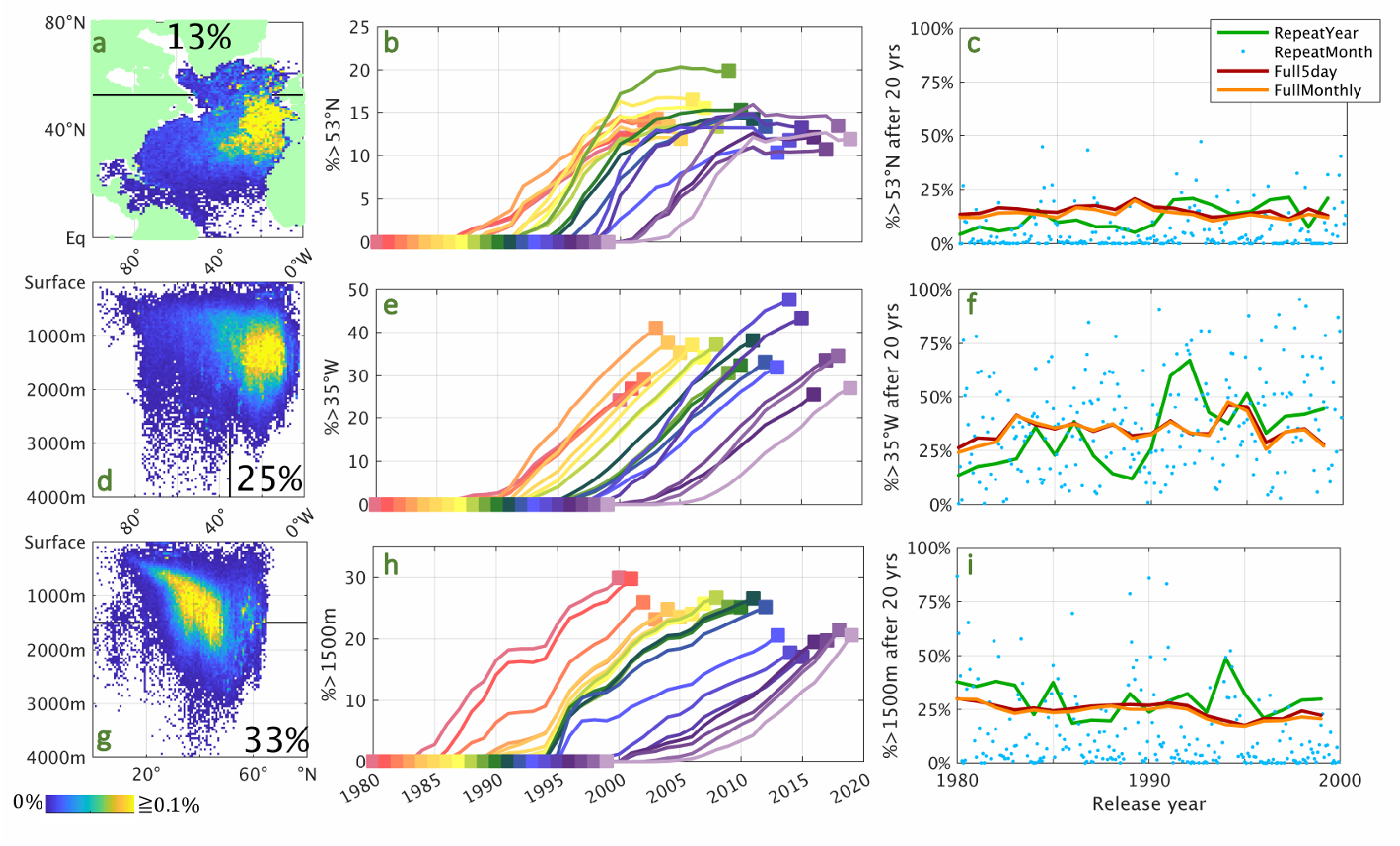}
\caption{
Spreading of the MOW in the North Atlantic.
(a,d,g) Probability density plots of MOW particles released in FullMonthly data from January 1992 to December 1992 once a month, measured 20 years after initial release, in January 2012. 
(b,e,h) show the percentage of particles beyond 53°N, 35°W, and 1500m, respectively, per year, where each color signifies a different release year, for FullMonthly data. Note the difference in the y-axis between subplots.
(c,f,i) show the overall percentage of particles crossing the benchmarks per release year, for all data types. The RepeatMonth experiment (blue) has a data point for every month repeated.}
\label{fig:spreading}
\end{figure*}

For each experiment, the particles are released once a month from the MOW section at the Gulf of Cadiz at $7^\circ W$. This section is chosen since it is fully inside the Gulf of Cadiz, westwards enough that it is beyond the major MOW sinking plume and eastwards enough that the velocity field is still clearly separated to an eastbound flow and a distinct westbound channel, see Fig.  \ref{fig:MOWtongue}, \ref{fig:whoisMOW}.
At this section, $7^\circ W$ between $34-37^\circ N$, the salinity and depth threshold for defining the MOW is determined by the depth and salinity values at which the maximum salinity per depth is at its minimum, since any further rise in the salinity is attributed to the MOW ((Fig. \ref{fig:whoisMOW}(a,b,c)). Finally, the MOW area is defined as the intersection between points with depth and salinity higher than the threshold values and points with a westbound velocity. In Fig. \ref{fig:whoisMOW}(d), the overall MOW transport in Sverdrup according to this definition is shown per month for the FullMonthly and Full5day datasets, showing an outflow of around 1 Sv with an interannual variability of up to 1.5 Sv.
Every month, we release between 1500-15000 virtual particles, such that each particle carries $10^{-4}$ Sv of water, distributed randomly on the MOW section at $7^\circ W$ as defined above.
The virtual particles are tracked in their corresponding velocity field for 20 years, for the four different oceanic models we consider.

 \begin{figure*}[t]
 \noindent\includegraphics[width=\textwidth]{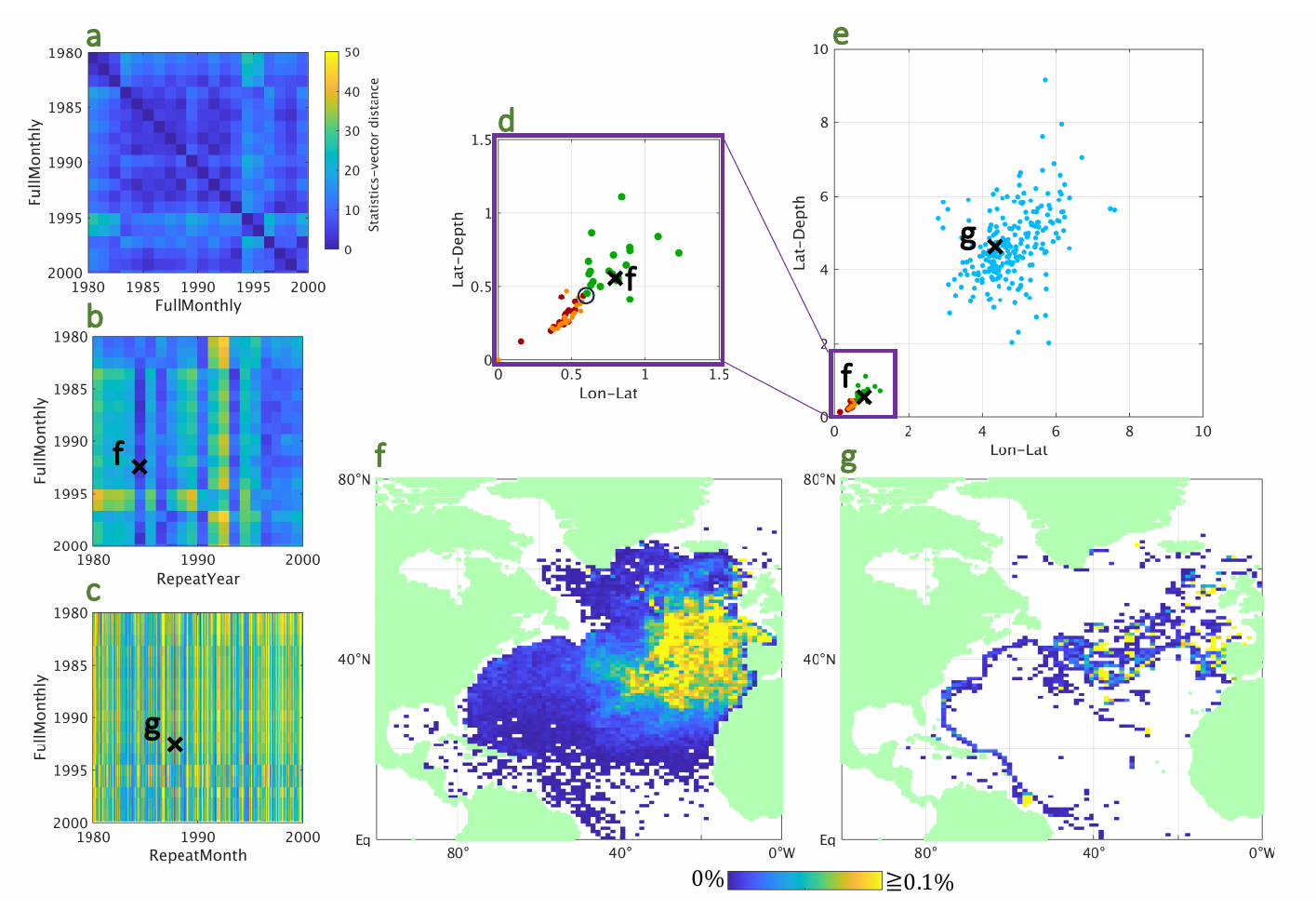}
\caption{
Mixing and statistics comparison.
(a,b,c) For each release year 1980-2020, the spreading statistics vector is defined as (\% north spreading after 20 years, \% west spreading after 20 years, \% sinking spreading after 20 years), and the pairwise Euclidean distances between each two releases are plotted, for FullMonthly from FullMonthly (symmetric matrix), RepeatYear, and RepeatMonth.
(d,e) show the KLD distance of spreading distributions from fM\_1992 (the year with statistics closest to their averages). The color scheme is the same as in Fig. \ref{fig:spreading}(c).
(f,g) show the longitude-latitude density plots of the RepeatYear (f) and RepeatMonth (g) releases with the smallest statistics vector distances, as marked by black x's on (b,c,d,e).
The black circle in (d) marks the RepeatYear experiment that has the smallest KLD distance from the origin.}
\label{fig:mixing}
\end{figure*}

\subsection*{Results}

\subsubsection*{Spreading and mixing of the different oceanic models}

To study the spreading of the MOW in the North Atlantic, we calculate the density of the final locations of MOW particles released over a single year and measured 20 years after the first release time (Fig.~\ref{fig:spreading}). We measure the percentage of particles northwards of 53°N, entering the sub-polar gyre region; westwards of 35°W, extending beyond the mid-Atlantic ridge; and deeper than 1500m, joining the NADW. In the Full5day and FullMonthly experiments, that are practically indistinguishable from each other, $>30\%$ of particles venture far enough to reach (at least) one of these regions. Specifically, an average of 13.5\% and up to 20\% of MOW particles reach the SPG region after approximately 15 years, at which point the entrance into and flushing out of the region reaches a balance, see a plateau in  Fig. \ref{fig:spreading}(b).
While differing in specifics, a similar qualitative picture is obtained for repearYear.
The SODA velocity fields do not resolve convective processes, and their vertical velocity is calculated diagnostically \cite{delworth2012simulated};
nevertheless, we measure a significant amount of approximately 35\% that sink under 1500 meters to join the NADW. The continuously positive slope of the curves in Fig. \ref{fig:spreading}(h) implies that continuing the trajectories beyond 20 years would raise this percentage even more.

The steady RepeatMonth runs, qualitatively different than the other  experiments, exhibit significant fluctuations in their northbound transport. Over 80\% of releases allow less than 5\% of particles to cross 53°N, while certain months exhibit a massive northbound transport volume of over 25\%.
Specific RepeatMonth releases do exhibit statistics that are close to those of the full and RepeatYear runs, however, they do not mix well (Fig. \ref{fig:mixing}), despite the steady velocity fields containing the same energy as the full runs.
To evaluate the degree of mixing, we use the symmetric Kullback-Leibler divergence (KLD)  \cite{seghouane2007aic} to calculate the distance between the spreading distributions. It is clear that the RepeatMonth runs are clustered far away and exhibit a much more limited spreading and mixing of the tracers. While Full5day and FullMonthly are again indistinguishable, the RepeatYear experiments are clustered somewhat separately. Nevertheless, there are RepeatYear runs that are indistinguishable from the full data, e.g. see the run marked by a circle in Fig. \ref{fig:mixing}(d). This suggests that a yearly-periodic velocity field, chosen with care, can provide an excellent imitation of the full dynamics, despite its lack of interannual variability.

 \begin{figure*}[t]
 \noindent\includegraphics[width=\textwidth]{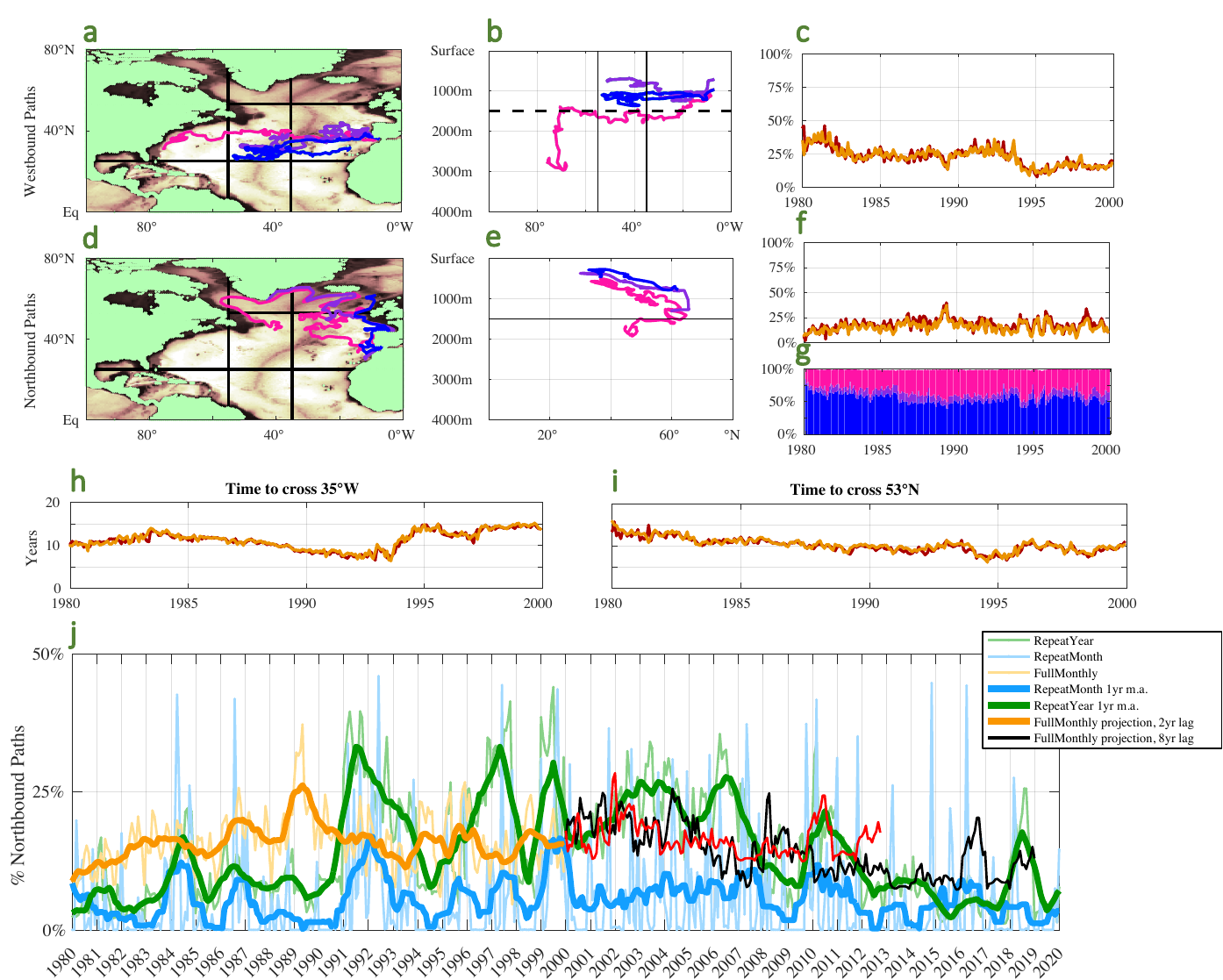}
\caption{
Typical pathways of the MOW.
(a-b) / (d-e) are typical examples of westbound / northbound trajectories, defined as pathways that exit the release box through its western / northern boundary, crossing the 35°W / 53°N mark, respectively. (c) / (f) show the percentage of particles that take the westbound / northbound path out of all the particles released in a single month, plotted as a function of the release month; the legend is the same as in Fig.~\ref{fig:spreading}. 
(g) shows the percentage of northbound particles that either stayed in the northeast box after exiting the initial box (blue), moved on to the north-center box (purple), or continued to the northwest box (pink).
(h) / (i) show the average amount of years it took the particles released at a given date to cross, respectively, 35°W / 53°N.
(j) 
Similar to (f), the percentage of particles that take a direct northbound path for FullMonthly, RepeatMonth, and RepeatYear, out of all particles released at a given date. 12-month moving averages (m.a.) are marked in bold. 
The black (red) prediction of the FullMonthly northbound percentage for 2000-2018 (2000-2012) is based on the 2-year (8-year) lag correlation between FullMonthly and RepeatYear; see App. C.}
\label{fig:typicaltraj}
\end{figure*}

\subsubsection*{The northward trajectories}

The horizontal dynamics of the Lagrangian trajectories divide into three distinct groups (Fig.~\ref{fig:typicaltraj}):
1. An average of 61\% of MOW trajectories are stationary, defined as contained in the box 25°N-53°N,  5°W-35°W throughout their 20 years of evolution;
2. 23\% of MOW trajectories take a direct westbound path, taking on average 11.3 years to cross the mid-Atlantic ridge at 35°W;
3. 16\% of trajectories follow a direct northbound path along the eastern boundary of the North Atlantic, taking on average 10.4 years to cross the 53°N mark.
A negligible amount ($<1\%$) of trajectories also exit the box from the south.
Out of the northbound particles, most particles (55\%) enter the northeastern box and stay there, whereas 34\% continue into the subpolar gyre, reaching the northwestern box.
Again, the FullMonthly and the Full5day trajectories are practically indistinguishable.


 \begin{figure*}[t]
 \noindent\includegraphics[width=\textwidth]{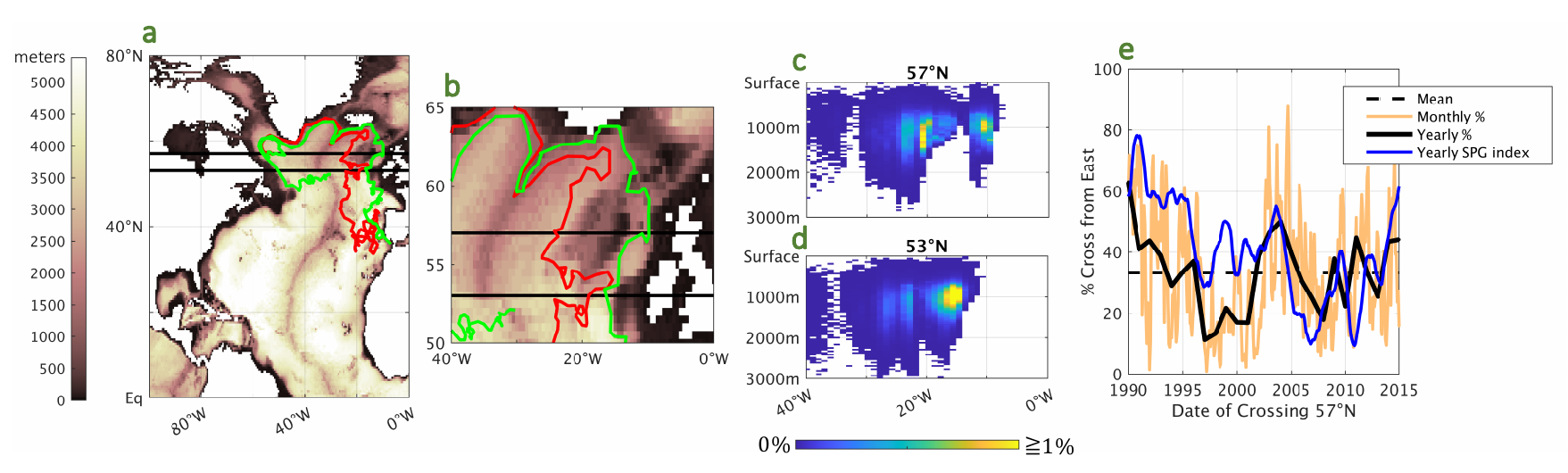}
\caption{
(a,b) show two typical FullMonthly trajectories that cross 57°N, via either an eastern path (green) or a western path (red) around the Rockall Ridge, both continuing into the sub-polar gyre.
(c,d) shows density sections at, respectively, 57°N and 53°N, of all trajectories from the FullMonthly releases that cross these sections.
(e) Orange line - the percentage of particles that cross from the east passage out of all particles that cross 57°N, vs. the date at which they cross this section. Black line - its yearly moving average.
The blue line is the yearly moving average of the normalized SPG index, see App. B.}
\label{fig:wherecross}
\end{figure*}

To study at which longitudes the MOW trajectories cross into the SPG region of the North Atlantic, we combine all FullMonthly trajectories that cross the 53°N line, no matter through which pathway, and calculate the density plot on this latitude.
A single window at 18°W and at a depth of approximately 1000 meters provides a subsurface pathway for the vast majority of northbound trajectories, see Fig.~\ref{fig:wherecross}. The window is situated at the southern border of Rockall Ridge, which has a depth of less than 1000 meters.
It is a saddle-point of the velocity field, from which northbound trajectories separate into two paths, one passing from the east, through the Rockall Trough, at 900 meters,  as seen in \cite{burkholder2011middepth}; and one from the west of the ridge, at 20°N and around 1100 meters. 
An average of 35\% of trajectories choose the east over the west path; per date of crossing, the percentage of east-bound trajectories has a positive correlation of $R=0.32$ with the SPG index calculated from the SODA data directly (the calculation is shown in App. B), with a statistical significance of $\alpha<0.001$.

\subsubsection*{Chaotic advection and correlations}

To identify the mechanism that transports the MOW into the SPG region of the North Atlantic, we compare the percentage of trajectories that take a northbound path in the FullMonthly, RepeatYear, and RepeatMonth datasets, see Fig.~\ref{fig:typicaltraj}(j). 
While there is a statistically significant ($\alpha<0.001$) correlation of $R=0.26$ between RepeatYear and RepeatMonth, the RepeatYear northbound trajectory statistics are systematically higher than those from the RepeatMonth dataset, with 14\% of RepeatYear trajectories and only 6\% of RepeatMonth trajectories taking a direct northbound route, as defined in the previous section.
An even stronger result emerges from the FullMonthly data, of which 16\% of trajectories take a direct northbound route.
This implies that for most months, the main mechanism that transports particles from the Gulf of Cadiz to the sub-polar gyre is time-dependent chaotic advection, and not a steady, direct northbound current.
The FullMonthly northbound trajectory statistics, out of which each datapoint contains data from 20 years ahead of the release date, exhibit a statistically significant ($\alpha<0.001$) high correlation with both the 8-year-lagged RepeatYear northbound data ($R= 0.27$) and the 2-year-lagged RepeatYear data ($R = 0.28$); while a correlation with an 8-year lag is expected given the time it takes trajectories to cross the 53°N line, the 2-year-lag correlation is yet to be explained. 
We hypothesize that a dynamical bridge situated south of 53°N is typically reached after 2 years, and once a particle crosses it, it will have a better chance of reaching the north; exploration of this idea is left to future studies.
In any case, these correlations suggest some degree of predictability, as marked in Fig.~\ref{fig:typicaltraj}(j).
According to the 2-year correlation, a steady decrease in the northward transport of the MOW is expected in the next two decades, see details in App. C.

\subsection*{Conclusions/discussion}

In this work, we have created a thorough survey of the various Lagrangian pathways taken by the MOW over the past four decades. 
We have shown that in the course of 20 years, the MOW mixes into the entire North Atlantic basin, from 10°N-70°N, and between the coasts (Fig. \ref{fig:spreading}, \ref{fig:mixing}), with a greater concentration around the Gibraltar Straits, between 25-53°N, 10-35°W, and 500-1500 m. After 15 years, the northbound influx and outflux rates reach a balance with an average of 13\% of MOW particles situated in the SPG region, i.e. beyond 53°N, indicating that this is the timeframe required to study MOW northbound transport.
We identify a direct northbound path from the Gibraltar straits along the eastern section of the North Atlantic that leads beyond 53°N into the SPG region (Fig. \ref{fig:typicaltraj}). Through this path, an average of 16\% of MOW particles take a direct northbound path into the SPG region; in some release years, over 20\% and up to 35\% take the direct northbound path.

The 53°N line contains a relatively narrow window at the southern tip of the Rockall Ridge, between 14-17°W and  800-1200 m depth, through which the vast majority of northbound MOW particles cross into the SPG region (Fig. \ref{fig:wherecross}). At this point, they cross the ridge by one of two possible paths, around the east or around the west of the ridge. The two pathways converge after crossing the ridge. While most (65\% on average) of these particles take the western path, there is a positive correlation between the SPG index and the percentage of particles that take the eastern path at a given date of crossing, supporting the idea that a high SPG index correlates with an eastward extension of the SPG's eastern boundary that blocks particles from taking the western path around the ridge. However, we have not yet found evidence that a high SPG index correlates with a decreased overall transport of the MOW into the SPG region.

Over half of the direct northbound transport in the time-dependent experiments is a result of time-dependent chaotic advection, and not due to  a steady northbound current, as indicated by the  sharp decrease in direct northbound transport measured for the 3D streamlines of the snapshot monthly velocity fields of the RepeatMonth experiment  (Fig. \ref{fig:typicaltraj}(j)).
A statistically significant 2-year-lag correlation between the direct northbound transport in RepeatYear and FullMonthly indicates an expected decrease in the northbound transport of the MOW in the next two decades. 
Exploration of the dynamical origins of this correlation is left to future studies.

Throughout all the diagnostics we consider, the FullMonthly and Full5day statistics are practically indistinguishable. 
We expect that additional averaging, as done for example in climatological oceanic models (App. A),
 will begin to ruin the statistics at timescales longer than the typical lifetime of mesoscale eddies, 
around a few months. 
The yearly-periodic RepeatYear experiments differ from the FullMonthly experiments both in their statistics time series and in their averaged degree of mixing (Fig. \ref{fig:spreading}(c,f,i) and  Fig. \ref{fig:mixing}).
In light of the presence of a RepeatYear run (indicated as the circled run in Figure 3d) that closely resembles full runs, it can be determined that the necessity of interannual variability for replicating the observed MOW spreading is not a significant factor. Instead, an appropriately selected yearly-periodic flow can reproduce transport dynamics similar in quality and quantity to those observed in the complete dynamics.
On the other hand, the spreading of the MOW in all the steady RepeatMonth oceanic models (as well as in the steady climatological model, see App. A) differs significantly from all the spreading observed in all of the FullMonthly cases. 
Hence, we propose that temporal variability is essential for imitating the full flow. Finally, we suspect that the mesoscale eddies do play an important role in the MOW spreading: a climatological averaged time-dependent flow has low kinetic energy content and its MOW spreading appears to be much more restricted than the FullMonthly spreading (see App. A). 
Future studies are needed to identify the relative importance
of mesoscale eddies and the kinetic energy content of the velocity fields
 in mixing and spreading the MOW.
 Additionally, it will be interesting to explore, in this 3D context,
 the possible compensation of the mesoscale eddies and/or of the velocity field time dependence by an isotropic diffusion term.

\acknowledgments
OSK acknowledges the support of a research grant from the Yotam Project and the
Weizmann Institute Sustainability and Energy Research Initiative; and the support of the Séphora Berrebi
Scholarship in Mathematics. VRK and OSK acknowledge the support of the Israel Science Foundation,
Grant 787/22.
VRK also acknowledges the support of The Estrin Family Chair of Computer Science and Applied Mathematics. HG acknowledges the support of the Vigevani Research Project Prize.

\bibliography{ref} 

\appendix

 \begin{figure*}
 \noindent\includegraphics[width=0.8\textwidth]{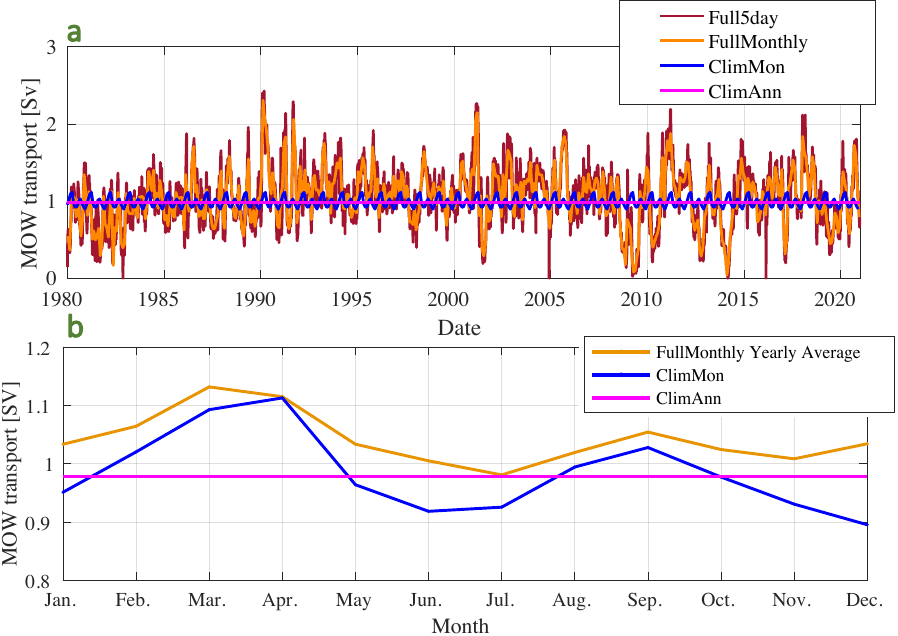}
\caption{
(a) The Mediterranean Outflow Water (MOW) transport in Sverdrup per release date, as defined in the main text, for the Full5day, FullMonthly, ClimMon, and ClimAnn experiments. (b) Yearly cycle of the MOW transport for ClimMon, ClimAnn, and FullMonthly.
}
\label{fig:appAMOWyes}
\end{figure*}

\section{Appendix A: Climatology}
\label{app:B}

The climatological velocity field is created by averaging over the entire database of SODA3.4.2 monthly velocity data, from 1980 to 2020.
To define the MOW section for these velocity fields, we create the corresponding climatological salinity fields from the SODA data and define the MOW in the same way as described in the main text, see Fig. \ref{fig:whoisMOW} and Fig. \ref{fig:appAMOWyes}.

We consider two different Lagrangian experiments:
\begin{itemize}
\item ClimMon: 12$\times$20 monthly releases in the yearly-periodic monthly climatological average of the data, denoted ClimMon\_\textit{\#year}\_\textit{\#month}. Since the monthly climatological data produces only 12 months of data, the different years are artificial: we create 20 different yearly releases with randomly different initial conditions between the years.
\item  ClimAnn: 12$\times$20 monthly releases in the steady annual climatological average of the data, denoted ClimAnn\_\textit{\#year}\_\textit{\#month}. Since the annual climatological data produces a steady velocity field, the different years and months are artificial: we create  12$\times$20 different monthly releases with randomly different initial conditions between the months and years.
\end{itemize}

 \begin{figure*}
 \noindent\includegraphics[width=0.8\textwidth]{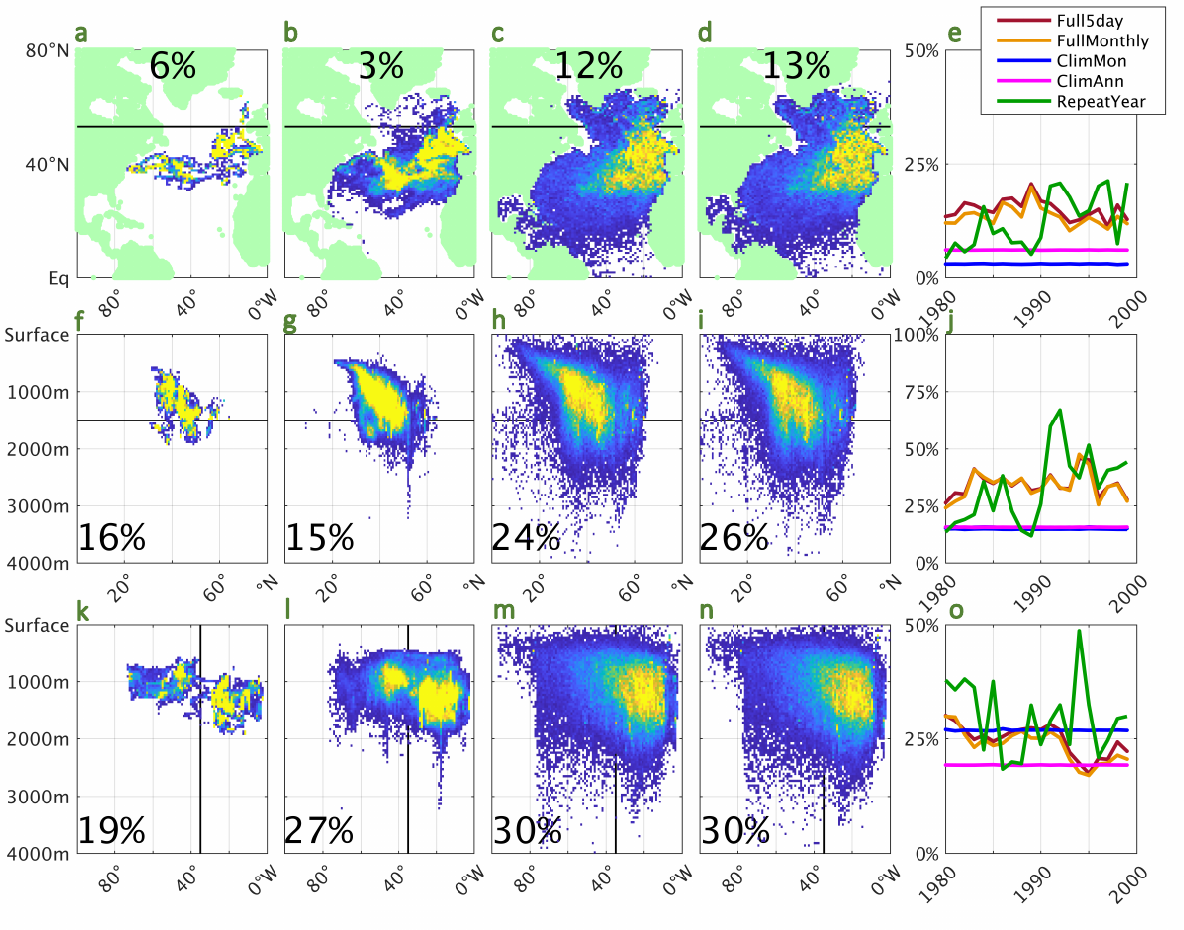}
\caption{
Spreading. (a-d,f-i,k-n) Probability density plots of MOW particles released from January 1980 to December 1980 once a month, measured 20 years after initial release, at January 2000. 
(a, f, k) ClimAnn; (b, g, l) ClimMon; (c, h, m) FullMonthly; (d, i, n) Full5day. (a-d) \% of particles ending up north of $53^\circ$N line is marked; (f-i) \% of particles below $1500$m is denoted; and in (k-n) \% westwards of $35^\circ$W is denoted. (e, j, o) show the percent of particles crossing, respectively,  $53^\circ$N, $1500$m, and $35°$W, for the five experiments noted in the legend.}
\label{fig:appAspreading}
\end{figure*}

\begin{figure*}
 \noindent\includegraphics[width=0.8\textwidth]{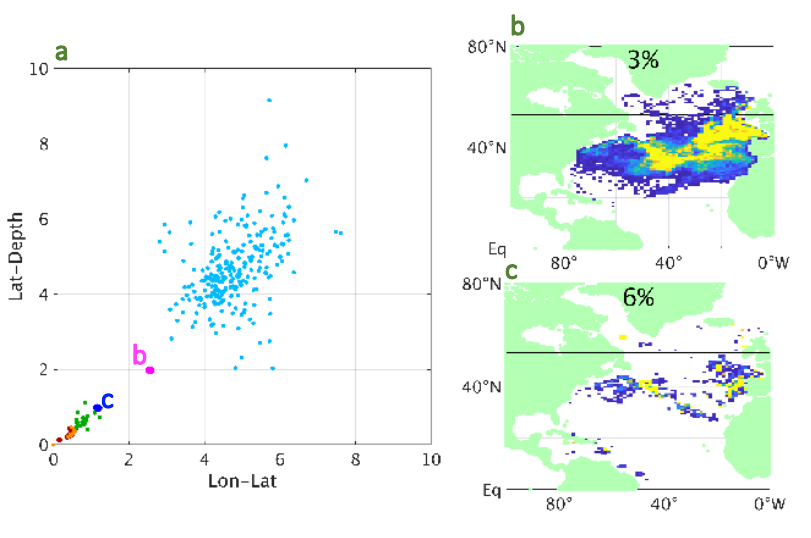}
\caption{
Mixing comparison.
(a) shows the KLD distance of spreading distributions as described in main text from fM\_1992 (the year with statistics closest to their averages). The color scheme is the same as in Fig. \ref{fig:spreading}(c) and Fig. \ref{fig:appAspreading}.
(b,c) show the longitude-latitude density plots of the ClimMon (b) and ClimAnn (c) releases.
}
\label{fig:appAKLDmixing}
\end{figure*}

In Fig. \ref{fig:appAspreading} we show the spreading of the MOW in the horizontal, zonal, and meridional sections, 20 years after the initial release, similar to Fig. \ref{fig:spreading} in the main text. We note that while the ClimAnn spreading is much more limited than the ClimMon spreading, it exhibits more northbound transport. On the contrary, the westbound transport is much more significant for the ClimMon experiments than the ClimAnn experiments. Both ClimAnn and ClimMon exhibit a similar percentage of sinking particles.
The limited spreading and mixing of the climatological experiments is reflected by the KLD measure (Fig. \ref{fig:appAKLDmixing}), which clusters the  ClimMon experiments near the far end of the RepeatYear experiments, and the ClimAnn experiments between the ClimMon and the RepeatMonth experiments. According to the spreading distributions, it seems that adding isotropic diffusion to the climatological data will not suffice to recreate realistic spreading and mixing, however since diffusion can help particles jump from cycle to cycle, further research is needed.

\begin{figure*}
 \noindent\includegraphics[width=0.8\textwidth]{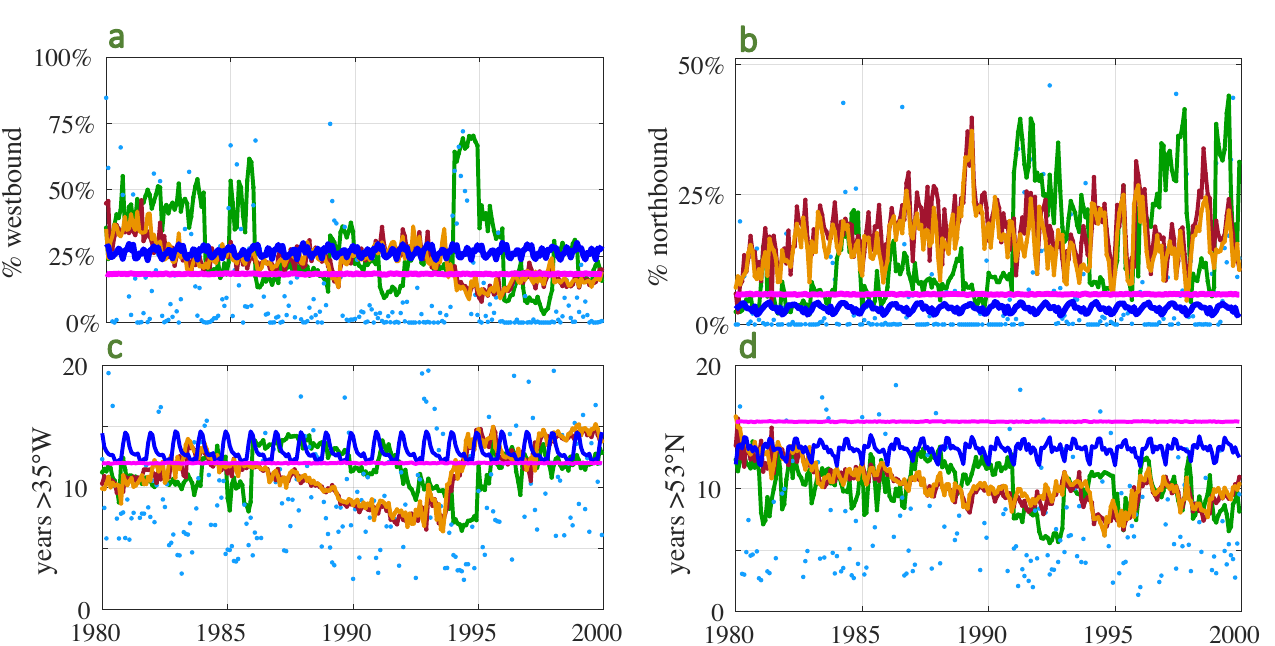}
\caption{
Typical pathways of the MOW, as defined in Fig. \ref{fig:typicaltraj} in the main text.
(a,b) show the percentage of particles that take each path out of all the particles released in a single month, plotted as a function of the release month; the legend is the same as in Fig.~\ref{fig:spreading} and Fig. \ref{fig:appAspreading}. 
(c) / (d) show the average amount of years it took the particles released at a given date to cross, respectively, 35°W / 53°N.
}
\label{fig:appAtypicaltraj}
\end{figure*}

The specific pathways and transport timescales taken by the climatological experiments are shown in Fig. \ref{fig:appAtypicaltraj}.
The percentage of particles that take a direct westbound route in the climatological experiments is close to the FullMonthly and RepeatYear averages. We note that there are more ClimMon direct westbound trajectories than ClimAnn; 
the dynamical and structural origins of this interesting phenomenon, in which the time-periodic transport leads more particles to the west while taking slightly longer to get there, are still a mystery.
On the contrary, there is a significantly lower percentage of northbound climatological trajectories - only 3\% of ClimMon and 6\% of ClimAnn trajectories take a direct northbound route. This indicates that the spatial averaging of the climatological velocity fields breaks up the northbound flow mechanism.
In this case, there is more northbound ClimAnn transport than ClimMon transport; however, the ClimMon northbound transport is quicker.
The spatial structure of this northbound mechanism, as well as the mechanism behind the difference in rate and percentage of transport between ClimMon and ClimAnn, is yet to be explored.

\begin{figure*}
 \noindent\includegraphics[width=0.8\textwidth]{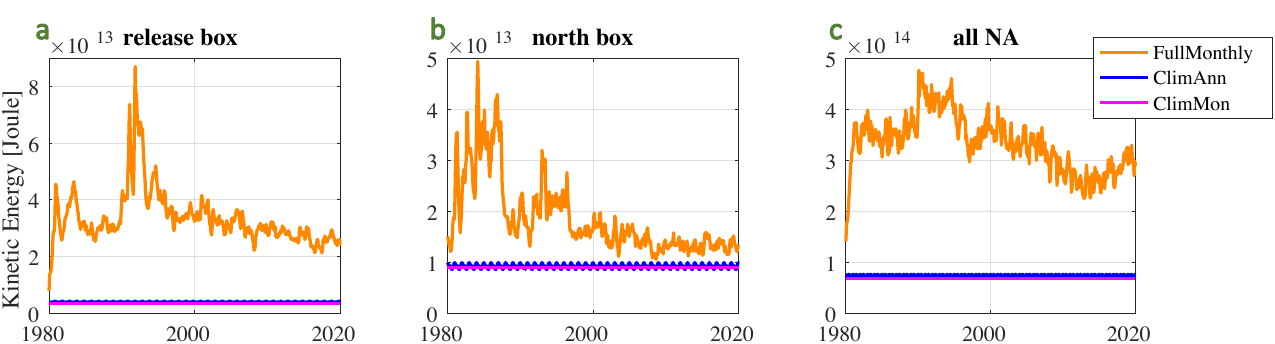}
\caption{
The total kinetic energy contained in the FullMonthly, ClimAnn, and ClimMon velocity fields in three different boxes: (a) the release box, between 25°N-53°N,  5°W-35°W;
(b) the north-east box, between 53°N-75°N, 5°W-35°W;
(c) the entire North Atlantic, between 10°N-75°N, 5°W-80°W.
The Full5day energy coincides with the FullMonthly energy and hence is not shown here.
}
\label{fig:appAkE}
\end{figure*}

The significant difference between the climatological and the full experiments can have several explanations:
\begin{itemize}
\item Lack of variability (for ClimAnn)
\item Lack of interannual variability (for ClimMon)
\item Reduction in small spatial scales due to averaging (for ClimMon and ClimAnn)
\item Reduction in eddy kinetic energy due to averaging (for ClimMon and ClimAnn), see Fig. \ref{fig:appAkE}
\end{itemize}
In the main text, we show that a steady flow with the same distribution of small spatial scales and the same kinetic energy as the full data (i.e. the RepeatMonth experiment) exhibits vastly different mixing and spreading than the full data; however, a well-chosen steady flow can recreate the gross statistics of the realistic transport.
We also show that a yearly-periodic flow with the same spatial scales and kinetic energy as the full data (i.e. the RepeatYear experiment) can recreate realistic mixing and spreading as well as transport statistics, indicating a reduced importance of interannual variability in establishing the transport.
However, we have not yet studied the relative contributions of small spatial scales and overall kinetic energy in creating realistic transport.
Specifically, would a yearly-periodic flow with reduced spatial scales, such as the ClimMon experiment, that has an enhanced amount of kinetic energy, achieved for example by raising all velocities by an appropriate factor, recreate realistic spreading, mixing, and statistics of transport? 
Since the kinetic energy content in the various experiments (Fig. \ref{fig:appAkE}) is not correlated with the KLD distance scattering (Fig. \ref{fig:appAKLDmixing}), we expect small scales to be imperative; however, this important question is yet to be answered.

\section{Appendix B: Calculating the Subpolar Gyre Index}
\label{app:A}

 \begin{figure*}
 \noindent\includegraphics[width=0.8\textwidth]{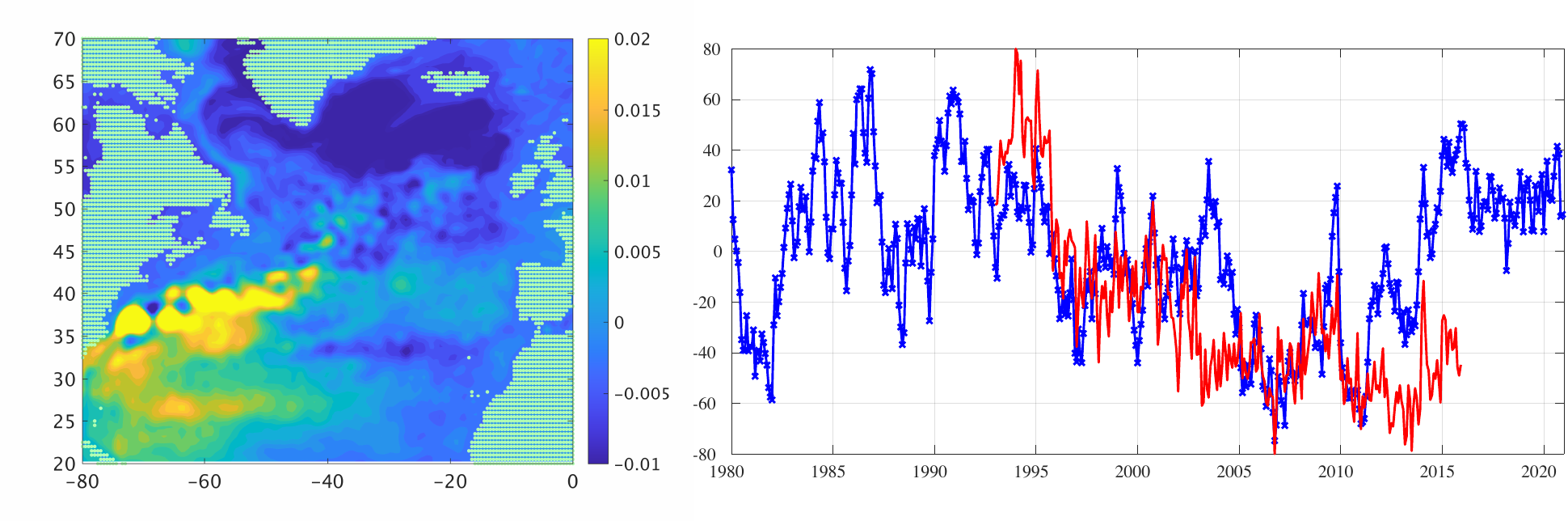}
\caption{
(left panel) EOF1.
(right panel) PC1 (blue) and Hakkinen \& Rhines SPG index, normalized (red). Correlation between blue and red: $R = 0.48$, statistical significance $>99.9\%$.
}
\label{fig:SPGindex}
\end{figure*}

We calculate the sub-polar gyre (SPG) index based on the North Atlantic sea-surface height (SSH) following the recipe described in \cite{hakkinen2004decline} and used in \cite{burkholder2011middepth}; see the overview on different options for insightful indices of the Atlantic sub-polar gyre in \cite{koul2020unraveling}.
We use the SSH data from the SODA3.4.2 database from 1980 to 2020.

We define a section: $0-80°W$, $20-70°N$, and look at the SSH in this section for 1980-2020 with a monthly resolution ($m$ time points).

We create the initial data matrix $X_1$: give each grid point in the section a number ($n$ grid points). Then the data matrix has a size $(n m)$:
$$
(X_1)_{i,j} = SSH(gridpoint=i)(time=j)
$$
Then, remove the climatological monthly average:
$$
(X_2)_{i,j} = (X_1)_{i,j} - \left<  (X_1)_{i}\right>_{month j}
$$
Then, standardize - zero mean on each column, divide by standard deviation:
$$
(X_3)_{i,j} = (X_2)_{i,j} - \left<  (X_1)_{j}\right>_{i=1}^n
$$
Remove overall mean:
$$
(X_4)_{i,j} = (X_3)_{i,j} - \left< X_3\right>_{i,j}
$$
Create the cross-correlation matrix:
$$
\Sigma = (X_4) \times (X_4)^T / (n-1)
$$

Diagonalize. Sort eigenvalues and vectors by size such that
$\lambda_1 \geq \lambda_2 \geq ... \geq \lambda_n$.

The variance percentage of the $i$'th mode is $\lambda_i/\sum_k{\lambda_k} * 100$. Thus, the first mode contains $7.1\%$ of the variance, the first 10 modes contain $31.9\%$ of the variance, and the first 25 modes contain $50\%$ of the variance.

The largest eigenvector $\hat{e}_1$ is called the Empirical Orthogonal Function (EOF1). It is of length n, and can be refolded into the initial geometry to see what the EOF1 looks like, see Fig. \ref{fig:SPGindex}(left panel). The physical structure of EOF1 signifies that the SPG index basically measures the SSH difference between the sub-tropical and the sub-polar gyres.

The first principle component, defined 
$\vec{P}_1 =  (\hat{e}_1)^T X_4$, is a vector of length $m$ that is defined as the SPG index. It is a time series of the change in the principle component of the SSH; see Fig. \ref{fig:SPGindex}(right panel). The y-axis is dimensionless.

The correlations between our SPG index and the \cite{hakkinen2004decline} SPG index, the NAO index, and the NAO-JFM index are, respectively, $R=0.48, 0.23, 0.51$, all with a statistical significance $>99.9\%$. Our SPG index also exhibits a statistically significant negative correlation with the NAO-JFM index of $R=-0.3$ with a time lag of 86 months (7.2 years) of the SPG index with respect to the NAO-JFM index.

\section{Appendix C: Calculating correlations and projections}
\label{app:calculatecorrelations}
In order to calculate the time-lagged correlation between two time-series $A$ and $B$, each of length $n$, we calculate their Pearson correlation coefficient using the Matlab corrcoef function, and calculate the corresponding p-value. 

Assume two vectors of different lengths, $A$ of length $n$ and $B$ of length $m$ where $m>n$.
If $A$ and the section of $B$ from $1$ to $n$, denoted $B(1:n)$, have a statistically significant correlation, it may be plausible to project future behavior of $A(n+1:m)$ through the values of $B(n+1:m)$. Assuming a linear correlation, one simply finds the best linear fit that goes through the function $A(1:n)$ as a function of $B(1:n)$, and uses this fit to find the values of $A(n+1:m)$ according to the values of $B(n+1:m)$.

Using this method, we calculate the projected percentage of FullMonthly direct northbound trajectories for particles released from 2000 to 2020 and tracked for 20 years, according to the 2-year and the 8-year lagged correlation between the FullMonthly and RepeatYear direct northbound pathways percentages. The projections are marked in Fig. \ref{fig:typicaltraj}(j).

\end{document}